# Surface roughness effects in a transonic axial flow compressor operating at near-stall conditions

Prashant B. Godse,[1] Harshal D. Akolekar,[1] and A. M. Pradeep[2]

[1)]*Department of Mechanical Engineering, Indian Institute of Technology Jodhpur, India*
[2)]*Department of Aerospace Engineering, Indian Institute of Technology Bombay, India*

(*Electronic mail: harshal.akolekar@iitj.ac.in)

**ABSTRACT**

Surface roughness is a major contributor to performance degradation in gas turbine engines. The fan and the compressor, as the first components in the engine's air path, are especially vulnerable to the effects of surface roughness. Debris ingestion, accumulation of grime, dust, or insect remnants, typically at low atmospheric conditions, over several cycles of operation are some major causes of surface roughness over the blade surfaces. The flow in compressor rotors is inherently highly complex. From the perspective of the component designers, it is thus important to study the effect of surface roughness on the performance and flow physics, especially at near-stall conditions. In this study, we examine the effect of surface roughness on flow physics such as shock-boundary layer interactions, tip and hub flow separations, the formation and changes in the critical points, and tip leakage vortices amongst other phenomena. Steady and unsteady Reynolds Averaged Navier Stokes (RANS) calculations are conducted at near-stall conditions for smooth and rough NASA (National Aeronautics and Space Administration) rotor 67 blades. Surface streamlines, Q-criterion, and entropy contours aid in analyzing the flow physics qualitatively and quantitatively. It is observed that from the onset of stall, to fully stalled conditions, the blockage varies from 21.7% to 59.6% from 90% span to the tip in the smooth case, and from 40.5% to 75.2% in the rough case. This significant blockage, caused by vortex breakdown and chaotic flow structures, leads to the onset of full rotor stall.

**Key Words**

Axial Flow Compressor, Surface Roughness, Stall, Computational Fluids Dynamics, Tip Leakage Vortex, NASA rotor 67

## I. INTRODUCTION

Industries that utilize gas turbine engines have consistently sought data on performance losses due to the in-service deterioration of components.[1–3] This need is especially urgent in the airline industry, where slight changes in performance can drastically affect specific fuel consumption and operational costs.[4–6] The fan and compressor, being the foremost components, are more prone to degradation from atmospheric particulates and runway debris. If the performance deterioration of these components can be accurately quantified, it will be possible to develop cost-effective repair strategies and determine optimal replacement intervals. Furthermore, understanding how blade deterioration occurs can help manufacturers and designers devise strategies to minimize long-term wear and tear. There are three major contributors to the deterioration of compressors. They are an increase in rotor tip clearance, changes in airfoil shape, and changes in the airfoil surface quality.[7] Of these three, only the degradation in performance due to airfoil surface quality is recoverable by compressor washing. The associated flow physics with surface roughness is thus a crucial parameter to understand for blade designers.

Compressors have different physical flow mechanisms occurring in them which lead to losses. Tip leakage vortices (TLV) and flows, shockwaves in compressors along with flow mixing, and shock-boundary layer interactions are among the prominent loss sources in compressors.[8] These secondary flows are governed by strong adverse pressure gradients (APG), shockwaves, endwall boundary layers, and the resulting aerodynamic effects.[9,10] Rotors and stators usually have different dominant flow mechanisms. The rotor often experiences rotating stall, marked by the development of rotating secondary flow structures. These structures cause flow blockage (a reduction in effective flow area due to phenomena such as boundary layer growth, flow separation, or the presence of vortices) and pressure fluctuations.[11,12] The stator is dominated by end-wall corner separations.[13] These secondary flows can lead to increased aerodynamic instabilities, flow distortion, and, ultimately, compressor stalling.[14,15] Thus, it is important to understand the unsteady secondary flow mechanisms for both smooth blades and blades having various degrees of roughness to enhance compressor aerodynamic stability and efficiency. Numerous researchers have tried to understand the origin and unsteady nature of secondary flows and to develop strategies to mitigate them.

The end wall regions of compressors are the most critical and least explored areas in the entire stage.[16] The flow in the end wall regions accounts for about a third of the total losses.[8] Casing and hub endwalls are two regions of the passage where most of the secondary losses are concentrated. TLV formation is one of the prominent vortex structures observed near the casing end wall region. Previous studies have shown that the existence of the TLV causes a range of flow phenomena, including TLV breakdown,[17,18] TLV fluctuations,[19,20] TLV/freestream interactions,[21] and TLV/shock interactions.[22] All of these cause substantial flow oscillations in the tip region. Yamada et al.[23] reported that the interaction of the lead-ing edge (LE) shockwave with the TLV leads to the latter's breakdown at near-stall conditions. The substantial blockage effect generated by the vortex breakdown results in a compressor stall. Kumar et al.[24] numerically investigated the evolution of unsteady secondary flow structures near the onset of stall in a tip-critical axial flow compressor stage. Babu et al.[25] studied the transient nature of secondary vortices in an axial compressor stage with a tandem rotor using unsteady Reynolds Averaged Navier Stokes (RANS) calculations.

The formation of secondary flow in the rotor is influenced by various factors, including tip leakage flow (TLF), which occurs due to rotor clearance and can worsen with tip shock. Storer et al.[26] conducted a comprehensive analysis of TLF, describing it as an inviscid phenomenon associated with blade loading. Dale et al.[27] further studied TLF, analyzing its formation, trajectory, and radial extent, and validated their findings with experimental data. Du et al.[28] highlighted the interface between TLF and core flow, and showed how each component affected stall inception differently. Zhang et al.[29] investigated changes in TLF during rotating stall inception, noting the unsteady breakdown of TLF and its spillage-inducing spike-type rotating stall. Researchers like Xiao et al.[18] utilized spectral proper orthogonal decomposition to study various flow structures within TLF, offering insights into their impact on aerodynamic and aero-elastic performance. Efforts have been made to mitigate TLF effects, including tip clearance adjustments and exploration of passive casing treatments. The research provided insights into how different flow structures within TLF impact both aerodynamic and aero-elastic performance. Zhang et al.[29] conducted numerical simulations, varying tip clearance, and discovered that reducing tip clearance to half of the design clearance enhanced efficiency by 0.2% and pressure ratio by 3.5%. However, this reduction also heightened TLF, potentially leading to rotating stall. Du et al.[28] conducted experimental investigations on TLF with varying gaps. They observed stable growth of TLF, which gradually transitioned into instability, leading to turbulent mixing with the core flow. This instability was closely linked to the evolution of flow blockage. Researchers have explored efforts to mitigate TLF effects. Jichao et al.[30] proposed the use of tip air injection, while Kumar et al.[31] implemented recirculation in the tip region. Additionally, passive casing treatments have been studied by Yan et al.,[32] and Alone et al.,[33,34] both aiming to minimize the adverse effects of TLF. Additionally, studies by Lei et al.,[35] Auchoybur et al.,[36] and Bailie et al.[37] focused on corner separation phenomena, providing criteria and modeling comparisons to predict and understand 3D hub corner separation in axial compressors. Techniques such as boundary layer suction slots and innovative slot configurations have shown promise in reducing flow separation. Li et al.[38] used large eddy simulations to explore the effects of laminar and turbulent inlet boundary layers on hub corner separation.

The interaction between the shock boundary layer is a major concern in transonic compressors. According to Hah et al.,[39] this interaction can lead to flow separation over 40% of the blade span. Schobeiri[40] introduced a shock loss model that effectively predicts the shock position, Mach number, and the total pressure losses induced by shocks. Furthermore,

Liu et al.[41] examined the impact of shockwaves on a transonic contra-rotating compressor stage under different operating conditions. The investigation revealed significant differences in shockwave structures between the two rotors under varying conditions, with the downstream rotor having more complex structures and higher losses. Biollo et al.[42] adjusted NASA (National Aeronautics and Space Administration) rotor 37's stacking line to align blade curvature with rotation, thereby enhancing efficiency, shifting the shock downstream, and reducing secondary flows from shock/boundary layer interaction.

Roughness plays a key role in affecting losses on compressors. Suder et al.[43] studied the effect of surface roughness and thickness on compressor configurations at various speeds. They found that the surface coating caused an increase in the boundary layer thicknesses and its interaction with the rotor passage shock caused an increased blockage and subsequent decrease in aerodynamic performance. Roberts et al.[44] studied the effects of equivalent sand grain roughness values up to 2 $\mu m$ on the NASA rotor 67 experimentally. When they ultra-polished the surface to approximately one-fourth of its initial roughness, the efficiency increased by 0.5% at design speed. Malhotra et al.[45] numerically investigated the effect of surface roughness due to uniform and non-uniform roughness for the NASA rotor 37. They found that roughness on the LE is the major contributor to loss compared to the trailing edge (TE). The roughness on the shroud is a greater contributor to the loss than the roughness on the hub.

It is easily inferred from the literature that understanding the secondary flow structures in transonic compressors with and without roughness is of utmost importance. There is a lack of numerical studies on the surface roughness effects of transonic compressors, and there is immense potential to gain a deeper understanding of flow physics, especially at near-stall conditions. Therefore, in this study, we attempt to investigate the effect of various roughness levels on the NASA rotor 67 especially close to the stall point. Firstly, a detailed validation is conducted against experiments using RANS for smooth and rough blades with an equivalent sand grain roughness[46] of 2 $\mu m$ and 30 $\mu m$. The flow characteristics of the above cases are discussed near stall. A detailed URANS study then follows, in which the flow physics and secondary structure formation of the smooth and 30 $\mu m$ rough NASA rotor 67 blades are discussed in detail from the onset of stall to fully stalled conditions. An equivalent sand grain roughness of 30 $\mu m$ is used as it represents the average value observed in a compressor first-stage blade after 20,000 operating cycles.[45] Thus, the results presented in this paper correspond to an in-service aerodynamic performance of the initial stage compressor rotor in a mid-size aircraft engine. The near-stall secondary flow structures evolving over the various time instances in the rotor passage for both smooth and rough cases are presented with the help of surface streamlines, entropy, and Q-criterion layover with Mach number contours. The blockage across the span for both cases is also quantified.

## II. NUMERICAL APPROACH

### A. Computational domain

The compressor stage considered for analysis in this study is the NASA rotor 67.[47] Only the rotor stage is simulated in this study. The rotor stage consists of 22 blades and has a design speed of 16043 rpm. The design tip speed is 429 m/s. The design mass flow rate is 33.25 kg/s and it has a total pressure ratio of 1.63. Figure 1 shows the geometry of NASA rotor 67 which has been used in this study. Stations 1 and 2 are measurement planes for inflow and outflow conditions. The inlet boundary is specified as a pressure inlet where the total temperature and pressure and the outlet boundary total pressures are prescribed. At the outlet, the static pressure is prescribed. The side walls have a rotational periodic boundary. The shroud is kept stationary and the hub with the blade fixed on it, is rotated. There is a gap of 1.016 mm between the shroud and the tip. The rotation direction is clockwise when seen from the outlet.

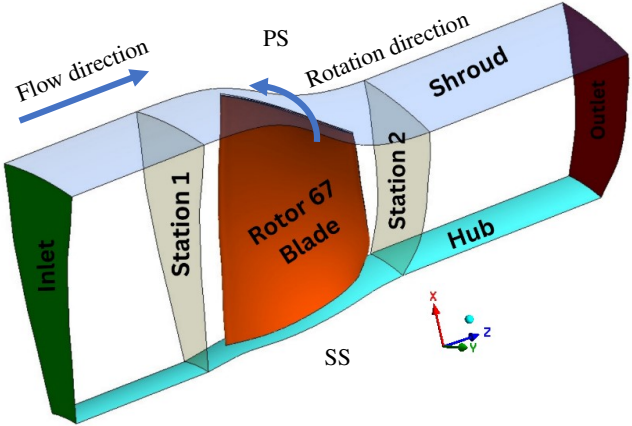

FIG. 1: Computational domain of the NASA rotor 67.

### B. Mesh generation

Mesh generation is carried out using ANSYS Turbogrid. An O-H mix topology is used to generate the structural hexahedral mesh. The H-grid topology is used in the passage area, while the O-grid topology is used in the blade region. Special attention is given to keeping the $y^+$ well below 1. Figure 2 shows the mesh generated for the stage with a zoomed view for the blade LE and TE regions.

### C. Solver details

Three-dimensional compressible RANS equations are used to do the calculations. The governing equations are solved using ANSYS CFX v2022R1,[48] a commercially available

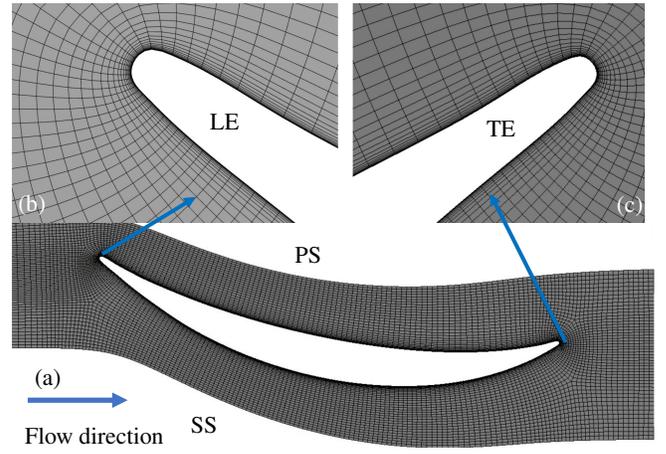

FIG. 2: (a) Structured hexahedral O-H mesh over the NASA rotor 67 blade. Mesh zoomed over the blade (b) LE and (c) TE

CFD solver. For steady calculations, a second-order high-resolution scheme with auto timescale is used for advection terms. The URANS calculations use a second-order backward Euler scheme for transient terms and for the advection terms second-order high-resolution scheme is used. The $k\omega$-SST[49] turbulence model is used for calculations as it can predict the complex turbulent flow quite well.[25,50] It contains two transport equations - one for the turbulence kinetic energy ($k$) and the other one for the specific dissipation rate ($\omega$) as shown in Eqs. (1) and (2).

$$\frac{\partial(\rho k)}{\partial t} + \frac{\partial(\rho u_i k)}{\partial x_i} = P_k - \beta^* \rho \omega k + \frac{\partial}{\partial x_i}\left[(\mu + \sigma_k \mu_t)\frac{\partial k}{\partial x_i}\right] + \rho G_k \tag{1}$$

$$\frac{\partial(\rho \omega)}{\partial t} + \frac{\partial(\rho u_i \omega)}{\partial x_i} = P_\omega + \beta \rho \frac{\omega^2}{k} + \frac{\partial}{\partial x_i}\left[(\mu + \sigma_\omega \mu_t)\frac{\partial \omega}{\partial x_i}\right] + \rho G_\omega \tag{2}$$

These equations are linked through a blending function ($\beta$), facilitating smooth adaptation of local flow fields. Here, $\mu$ represents dynamic viscosity, $\mu_t$ is the turbulent eddy viscosity, $G_k$ and $G_\omega$ are terms source terms for $k$ and $\omega$, and $\sigma_k$ and $\sigma_\omega$ are constants. $P_k$ and $P_\omega$ are production terms.

The details of the transient calculations are as follows. The passing period of a single blade is the time taken by the rotor blade to pass one pitch of the rotor blade. One passing period is divided into 20 physical time steps and 10 coefficients of loop iteration. Each time step corresponds to $1.699\times10^{-4}$s. The total time steps for ten revolutions is 4400, which is found to be adequate to achieve the limit cycle of the solution. The static pressure signal's time history is monitored, and the solution is considered converged when the pressure signal's periodicity becomes consistently repetitive.

### D. Mesh independence test

A mesh independence test is carried out for five meshes. The meshes have 0.35, 1, 2, 2.5, and 4 million elements respectively. Figure 3 showcases the mesh independence test with a plot of isentropic compression efficiency ($\eta$) against the mesh count in millions. The percentage variation in the efficiency was found to be 0.068% between the 2.5 million and 4 million meshes. Beyond the mesh size of 2.5 million, the variation in the efficiency is negligible and thus is an optimal mesh size for further analysis. The 2.5 million is thus used for further analysis.

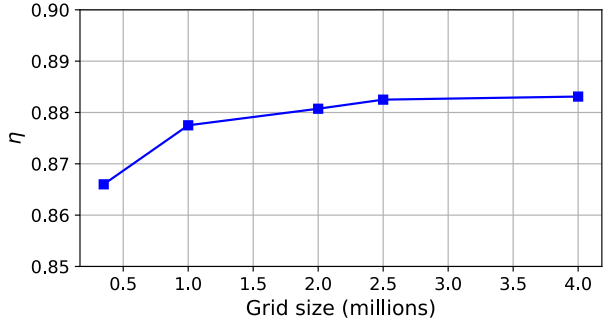

FIG. 3: Mesh independence test: isentropic compression efficiency ($\eta$) against the mesh count (in millions).

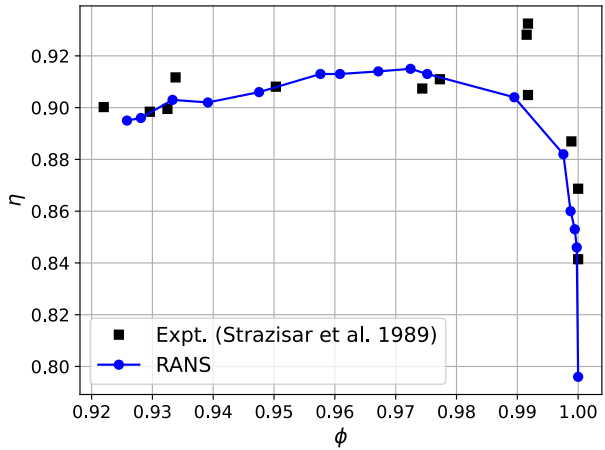

FIG. 4: Isentropic compression efficiency against the normalized mass flow rate for the smooth blade.

### E. Validation

Using the above mesh, steady-state analysis is carried out for both smooth and rough rotor blades. Experimental results from Strazisar et al.[47] are used for validation. Figure 4 shows the performance curve of isentropic compression efficiency against the normalized mass flow rate ($\phi$). Changing the back pressure varied the normalized mass flow rate for each RANS calculation. The numerical results match quite closely with the experimental results, with the largest deviation being 3.5% near the peak efficiency point for the experiments. The performance curve of the pressure ratio ($\Pi$) against the normalized mass flow rate is shown in Fig. 5. The pressure ratio from numerical results matches the experiment quite closely and has a maximum deviation of 4% at the near-stall condition. The last stable numerical solution is the near-stall condition. This would be the leftmost point from the RANS curve on the performance maps. These deviations are considered well within acceptable limits and we can conclude that the numerical results match both qualitatively and quantitatively with the experiments.

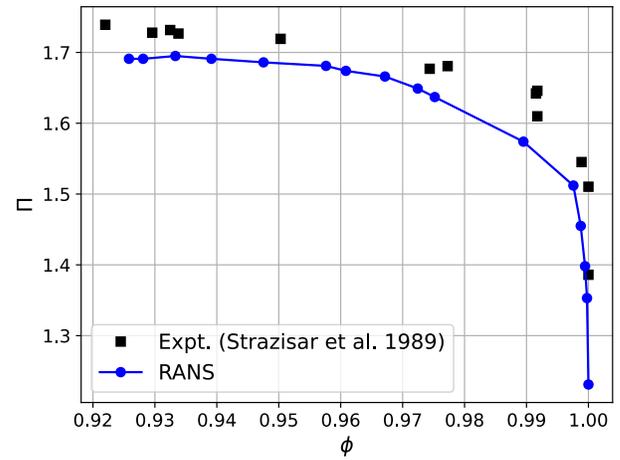

FIG. 5: Pressure ratio against the normalized mass flow rate for the smooth blade.

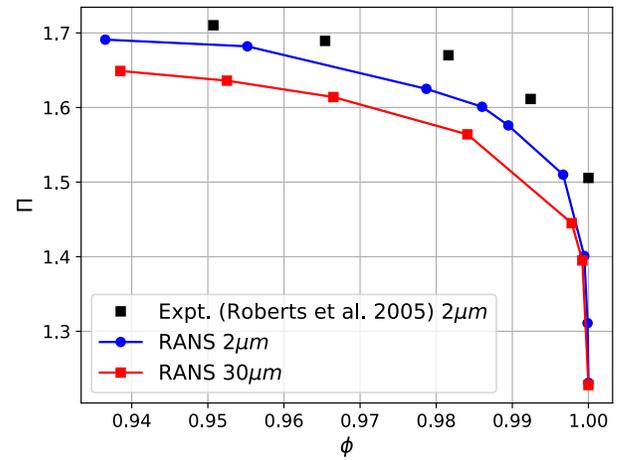

FIG. 6: Pressure ratio against the normalized mass flow rate for rough blades.

Validation was also conducted for a case with surface roughness. The roughness model available in ANSYS CFX[48] has been used in this study. In the numerical model, the equivalent sand grain roughness height is specified and the solver computes the effect of roughness on the flow field. The equiv-alent sand grain roughness is 3.1 times the RMS sand grain roughness.[46] For the roughness validation, the results of the rough NASA rotor 67 are compared with the experimental data of Roberts et. al[44] for an equivalent sand grain roughness of 2 $\mu m$, applied to both the pressure and suction sides of the blade.

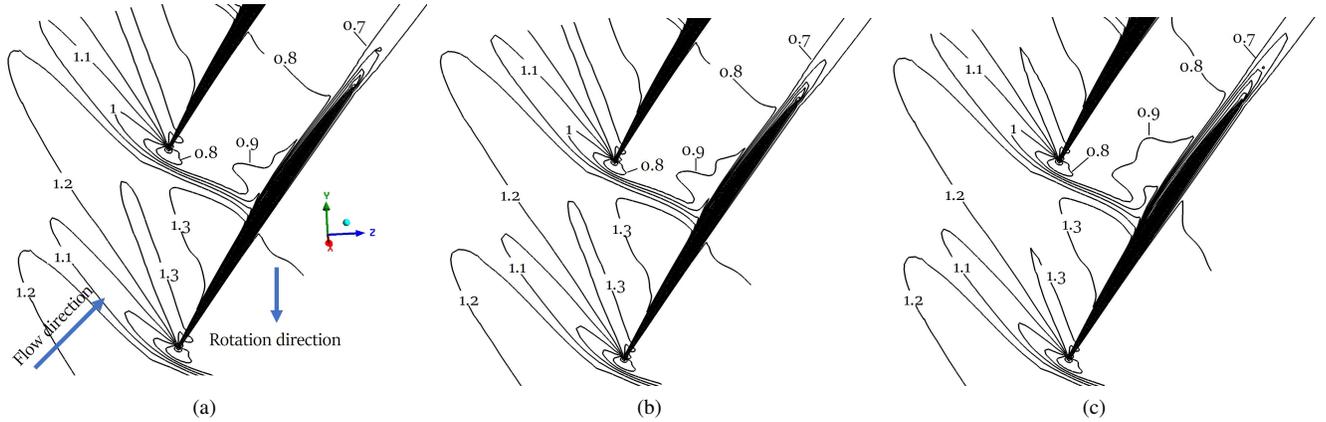

FIG. 7: Contours of relative Mach numbers at near-stall and 70% span from the hub for (a) smooth, (b) 2 $\mu m$, and (c) 30 $\mu m$ roughness cases.

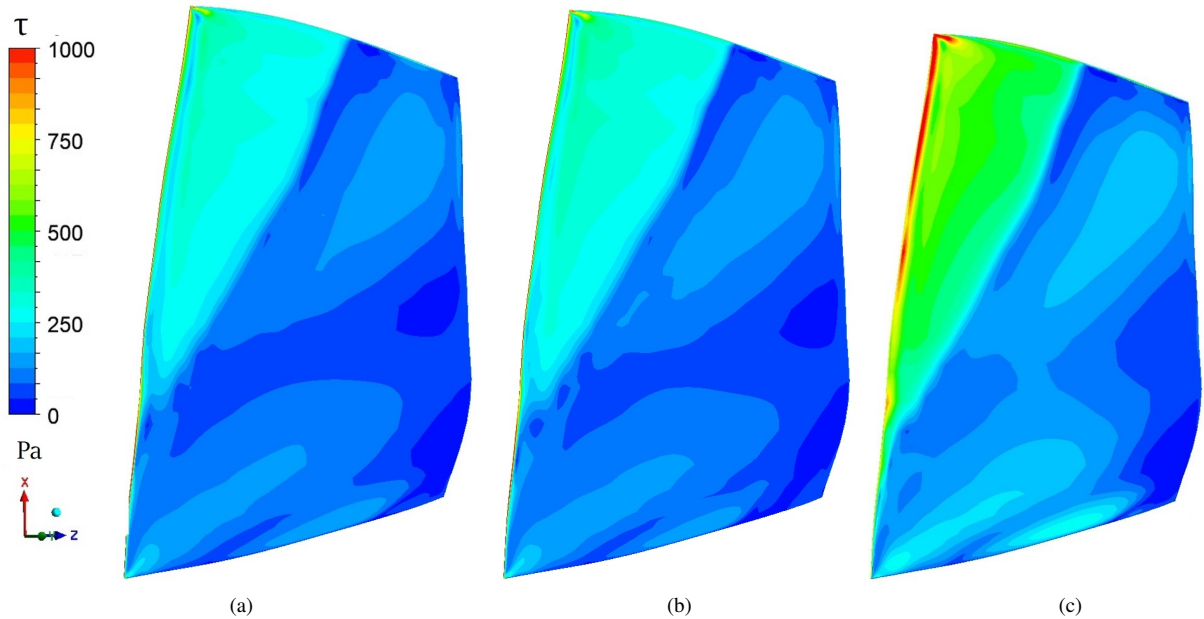

FIG. 8: Wall shear stress magnitude on the suction side for (a) smooth, (b) 2 $\mu m$, and (c) 30 $\mu m$ roughness cases.

Figure 6 shows the total pressure ratio against the normalized mass flow rate for the rough cases. The 2 $\mu m$ RANS calculations closely match the experiment and all points are within acceptable limits. The 2 $\mu m$ surface case has a maximum pressure ratio quite close to the smooth case (Fig. 5), and has a normalized mass flow rate of 0.935 near-stall, whereas the smooth case has a normalized mass flow rate of 0.925 near-stall. Since we would like to investigate the flow characteristics after 20,000 operating cycles, an additional line with an equivalent sand grain roughness of 30 $\mu m$ is also shown in Fig. 6. The stall point occurs slightly before the 2 $\mu m$ case. The maximum pressure ratio is also lower than the 2 $\mu m$ case, which is expected due to the increased roughness. From Fig. 7, the relative Mach number contours and the CFD's predicted shock structure for the smooth and 2 $\mu m$ rough blades match quite closely with their respective experimental data[44,47] at 70% span from the hub. Mach contours for the 30 $\mu m$ rough blade case are also shown.

## III. RESULTS AND DISCUSSIONS

In transonic rotors, the flow dynamics are quite complex due to several factors. As the flow approaches the rotor blade,

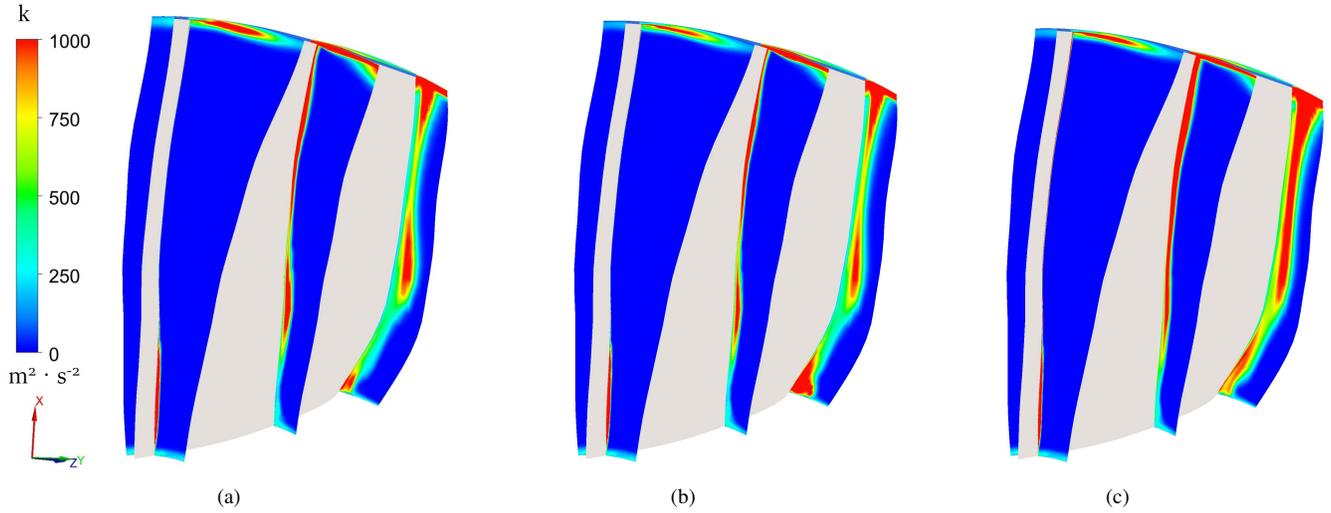

FIG. 9: TKE profiles for the (a) smooth, (b) 2 $\mu m$ and, (c) 30 $\mu m$ roughness cases at 0.1, 0.5 and 1 tip axial chord from the LE.

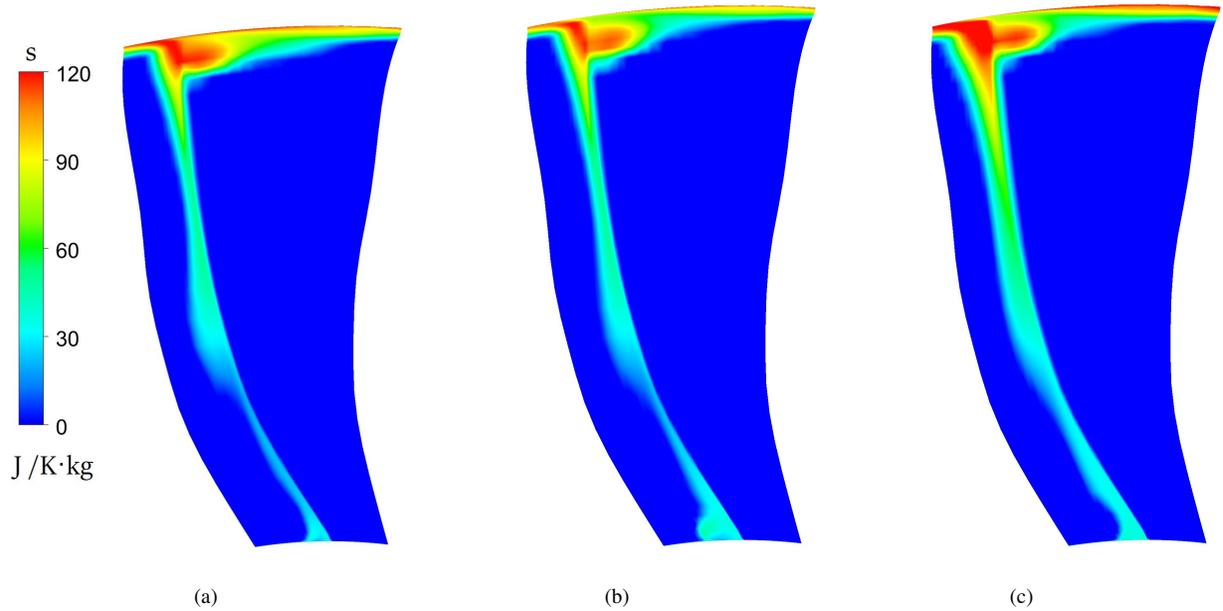

FIG. 10: Spanwise entropy at the TE surface for (a) smooth (b) 2 $\mu m$ and, (c) 30$\mu m$ roughness cases.

it accelerates over the initial part of the blade surface. This acceleration is influenced by the blade's shape and the relative motion between the blade and the incoming air. Since the tip speed is a function of the radius, the velocity at the blade tip can become very high. This can result in the local relative flow velocity at the tip reaching supersonic speeds, leading to the formation of shockwaves. The shock formation can significantly impact the blade's performance by causing local increases in pressure and temperature, as well as potential flow separation. The situation is further complicated by the presence of a three-dimensional pressure in the radial, tangential, and streamwise directions, creating a highly complex flow field. One of the largest sources of loss generation in a rotor section is the tip leakage flow. The above flow physics becomes more complex at near-stall conditions. This section discusses key flow physics for smooth and rough blades obtained from steady and unsteady RANS calculations at near-stall conditions.

## A. Flow Physics from RANS Near-Stall

The flow physics near-stall from steady RANS calculations for the smooth, the 2 $\mu m$ roughness and 30 $\mu m$ roughness cases are briefly discussed here. Figure 8 shows the wall shear stress magnitude ($\tau$) for the three cases on the suction side of the blade. The wall shear stress plots between the smooth case and the 2 $\mu m$ roughness cases are quite similar qualitatively and quantitatively. However, there is a marked increase in wall shear stress on both the suction and pressure sides (not shown here) for the 30 $\mu m$ roughness case due to higher roughness levels. The average wall shear stress across both sides of the blade is 309 Pa for the smooth case, 332 Pa (7.4% rise over the smooth case) for the 2 $\mu m$ roughness, and 395 Pa (27.8% rise over the smooth case) for the 30 $\mu m$ roughness case. In all three cases on the suction side, there are higher wall shear stress levels from the LE to the shock impingement location (approx. 60% tip axial chord for all cases). Downstream of the shock location, the wall shear stress levels drop drastically due to a reduction in velocity and thus reduced friction and wall shear for all three cases. On the suction side, a LE separation zone is observed near the hub for all three cases. The hub corner separation is also similar in size and extant for all cases. The wall shear stress on the pressure side of the 30 $\mu m$ roughness case is observed to be higher at the shroud LE than the hub LE due to higher velocities at the tip than the hub.

Figure 9 shows the turbulent kinetic energy (TKE) for the three cases along the entire span of the suction side at 10%, 50% tip axial chord, and at the TE. The 30 $\mu m$ roughness case has much higher TKE levels throughout the boundary layer in the spanwise direction from the LE to the TE. The increased TKE leads to higher mixing losses and greater blockage downstream of the blade leading to a decline in performance. The smooth case and the 2 $\mu m$ roughness case have very similar TKE plots, except near the hub portion of the TE. The 2 $\mu m$ roughness case shows marginally higher TKE near the TE as compared with the smooth case. In the 30 $\mu m$ roughness case, there are high TKE levels across the entire span, all the more in the tip region. Entropy generation is one of the best measures of loss[8] in a turbomachine. Figure 10 shows the entropy for all three cases along the span at the blade TE surface. The entropy at the TE surface would include the accumulation of all losses from the rotor passage and blade boundary layers. Higher span locations have increased losses due to a combination of high blade speeds, tip leakage flows, and shock-boundary layer interactions. The blade with the 30 $\mu m$ roughness shows the highest entropy across the span due to greater flow mixing resulting from the rough boundary layer. This is even more prominent in the tip region, where there is greater entropy generation due to stronger tip vortices. The smooth case and 2 $\mu m$ roughness cases have similar levels of entropy throughout the domain, with the 2 $\mu m$ roughness case having slightly higher losses in the tip region.

From the above images, there are numerous qualitative similarities in the flow physics between the smooth and 2 $\mu m$ roughness cases. This is especially true for the blade pressure and suction surfaces and across most of the span, except some portions of tip regions. Therefore, further URANS results are shown only for the smooth and the 30 $\mu m$ roughness cases (henceforward referred to as the 'rough case').

## B. Instantaneous Surface Streamlines using URANS

The complexity of the flow can be observed through the evolution of the instantaneous limiting streamlines. They illustrate how the flow is influenced by the blade geometry and the three-dimensional pressure gradient. The interaction between the shocks and the pressure gradients can result in intricate flow patterns, which are critical to understand for improving the rotor blade design and the overall performance. Surface streamlines on the suction side are plotted near the onset of stall for the smooth case in Figure 11. A time interval of $\Delta t = 0.51ms$ is used. $T_1$ indicates stall onset, whereas $T_6$ represents the fully stalled state. At $T_1$, the accelerating flow creates an LE shockwave which impinges at about 65% tip axial chord. This shock propagates radially downwards with reduced strength. The shock boundary layer interaction causes flow deflection as shown by the flow deflection line (DL) from the tip region to about 26% span. There is a LE separation zone (LESZ) up to around 26% of the span from the hub due to a local shock (secondary shock). The LESZ is encompassed by the attachment line (AL). The radial pressure difference, generated by the accelerating flow towards the tip and the diffusing flow near the hub, is further increased by the pressure rise from the secondary shock, causing the hub flow to drift radially upwards. The trajectory of the drifting flow (DF) is shown by the green line (DF), separating it from the shock-deflected flow.

There is a distinct hub corner separation zone (HCSZ) that begins near at 75% hub axial chord on the hub endwall and goes to 28% span on the hub TE, as shown by the hub separation (HS) line. There are no unstable points in the HCSZ at $T_1$ or $T_2$. At $T_2$ the flow deflection line originating from the tip region shifts slightly upstream. Overall, the streamline structure is quite similar to that of $T_1$. At $T_1$ and $T_2$ the streamlines on the pressure side (pressure side images are not shown here) do not show any instabilities. At $T_3$, the primary shock impingement is shifted to the extreme upstream near the blade LE, with flow beginning to separate as shown by the tip separation (TS) line. The hub drifted flow (DF) line divides the flow from the hub and the tip separation zone and shifts closer to the hub. The shock-deflected flow line (DL) merges with the hub drifted flow line around 43% span due to the formation of the tip LE flow separation. The DL line becomes less prominent at further time instances due to the increasing intensity of the tip flow separation. The LESZ is the same size as $T_2$ as is the hub corner separation zone. On the pressure side, there is a formation of a circulation zone about 40% tip axial chord and 97% span. The flow inside the hub corner separation has begun to grow unstable. There is a formation of a weak focus (FP$_1$) inside the HCSZ. A weak focal point has also begun to form on the hub surface (FP$_2$). At $T_4$, the tip LE separation starts from about 94% span. There is a developing focus near the tip LE (FP$_3$) and also the formation of a focus point (FP$_4$) around 17% of the tip axial chord. Due to the interaction of the separated flow from the tip LE, with the flow near the two foci, two saddle points are formed near the tip region (SP$_1$, SP$_2$). A nodal point (NP$_1$) is formed at about

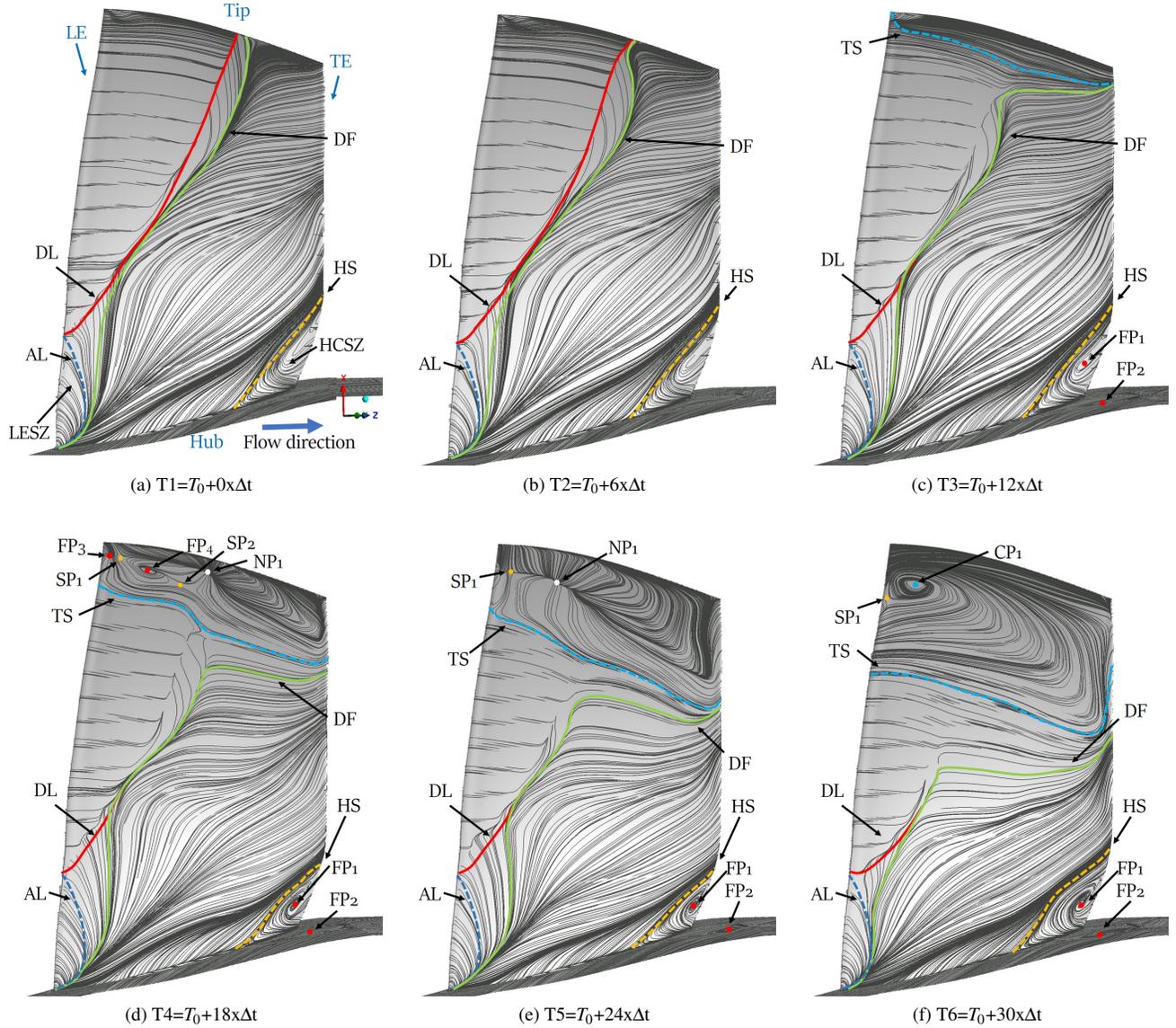

(a) T1=$T_0$+0x$\Delta$t

(b) T2=$T_0$+6x$\Delta$t

(c) T3=$T_0$+12x$\Delta$t

(d) T4=$T_0$+18x$\Delta$t

(e) T5=$T_0$+24x$\Delta$t

(f) T6=$T_0$+30x$\Delta$t

FIG. 11: Evolution of the suction surface streamlines at various time instances from the near-stall point for the smooth case.

50% of the tip axial chord due to the excessive instabilities in the flow. On the pressure side, the circulation zone from $T_4$ becomes a focus point at the same spanwise location and shifts slightly closer to the LE. There is a saddle point right near the LE, at 90% span on the pressure side. At $T_5$, the tip LE flow separation increases to 85% span, which increases the instabilities in the flow making the focus point (FP$_2$) vanish. The node (NP$_1$) becomes more unstable and moves upstream. The saddle point (SP$_1$) shifts slightly away from the tip. At $T_6$, the tip flow separation region on the suction side increases further in size, and a large circulation zone is formed which encompasses both the suction and pressure sides. The center of circulation (*CP*) for both these sides is at 90.5% span and 11% tip axial chord on the suction side. All saddle points and node points vanish on both sides of the blade, except for the saddle point (SP$_1$). The focus points in the hub corner region,

(FP$_1$) and (FP$_2$), which emerge at $T_3$, grow in intensity and thus make the flow more chaotic and unstable till $T_6$.

Similar streamlines are plotted on the suction side for the rough case (with 30 $\mu m$ roughness) in Figure 12. At $T_1$, the streamline configuration is similar to the smooth case. The attachment line (AL) line encompasses the LESZ. There is a hub corner separation which originates at 75% hub axial chord. Due to the shock impingement at 61% tip axial chord, there is a flow deflection (DL) line that originates from the tip to 24% span. The hub drifted flow line (DF) is also shown and extends from the LE of the hub to the tip region. At $T_2$, flow separation takes place, shown by the tip separation line (TS$_1$), which occurs earlier than the smooth case. For flows with just an APG, a rough surface, due to increased local turbulence, is likely to delay flow separation. However, in the present case, besides an APG, there are other complex flow features such as the TLF, shock wave, rotation of the blade, etc. These as well as the interactions between these different mechanisms cause the rather early onset of the flow separation as compared to the smooth case. This causes flow deflection and hub drifting flow lines to move towards the hub. Two focal points ($FP_1$ and $FP_2$) start developing inside the HCSZ, which grow in intensity till $T_5$. At $T_3$, unlike the smooth case, two separation regions begin to form near the tip region. One begins from the LE (92% span) and reattaches at 72% tip axial chord ($TS_1$). This is because the increased turbulence generated by the rough surface can energize the boundary layer, increasing its momentum near the wall. This enhanced momentum helps the flow overcome APGs, which leads to reattachment after the separation point. Right after reattachment, the flow separates again ($TS_2$) and reattaches at the TE resulting in a closed separation. Due to the interaction of the flow coming in from the hub and TE flow separation, a saddle point ($SP_1$) and an unstable node ($NP_1$) are formed. The LESZ and HCSZ remain almost unchanged as compared to $T_1$. At $T_4$, the flow separation line $TS_1$ reattaches at 60% tip axial chord, further upstream than at $T_3$. $TS_2$ develops into a bigger closed separation. There is a node ($NP_1$) in almost the same location as $T_3$. A node point ($NP_2$) emerges near the hub TE. At and beyond $T_4$, the radial extent of the separation is greater in the smooth case than in the rough case. The surface roughness keeps the boundary layer attached on a greater portion of the blade. At $T_5$, the flow becomes more unstable as the fully stalled condition approaches, and there is only one flow separation region at the tip, from the LE to the TE ($TS_1$). The node $NP_1$ migrates upstream with the corresponding saddle point $SP_1$. Another node ($NP_3$) forms near the tip LE. Saddle point ($SP_2$) appears between 50% axial chord and the TE, in the tip region. An unstable focus point ($FP_3$) forms near this saddle point. At $T_6$, the fully stalled condition, a large circulation zone, centered at $CP_1$ (28% tip axial chord, 94.5% span) emerges, that encompasses both the suction and pressure sides. There is a small LE separation zone that emerges near the LE of the tip. It is demarcated by the tip attachment (TA) line. Saddle point ($SP_1$) migrates near the tip LE. The HCSZ shrinks in size due to the increased flow separation at the tip region. However, it becomes stronger in intensity due to the formation of a strong circulation zone, centered at $CP_2$. The various flow separation lines and the emergence of a number of critical points and large circulation zones give a fair idea of the complex unsteady flow behavior as the rough rotor goes into stall.

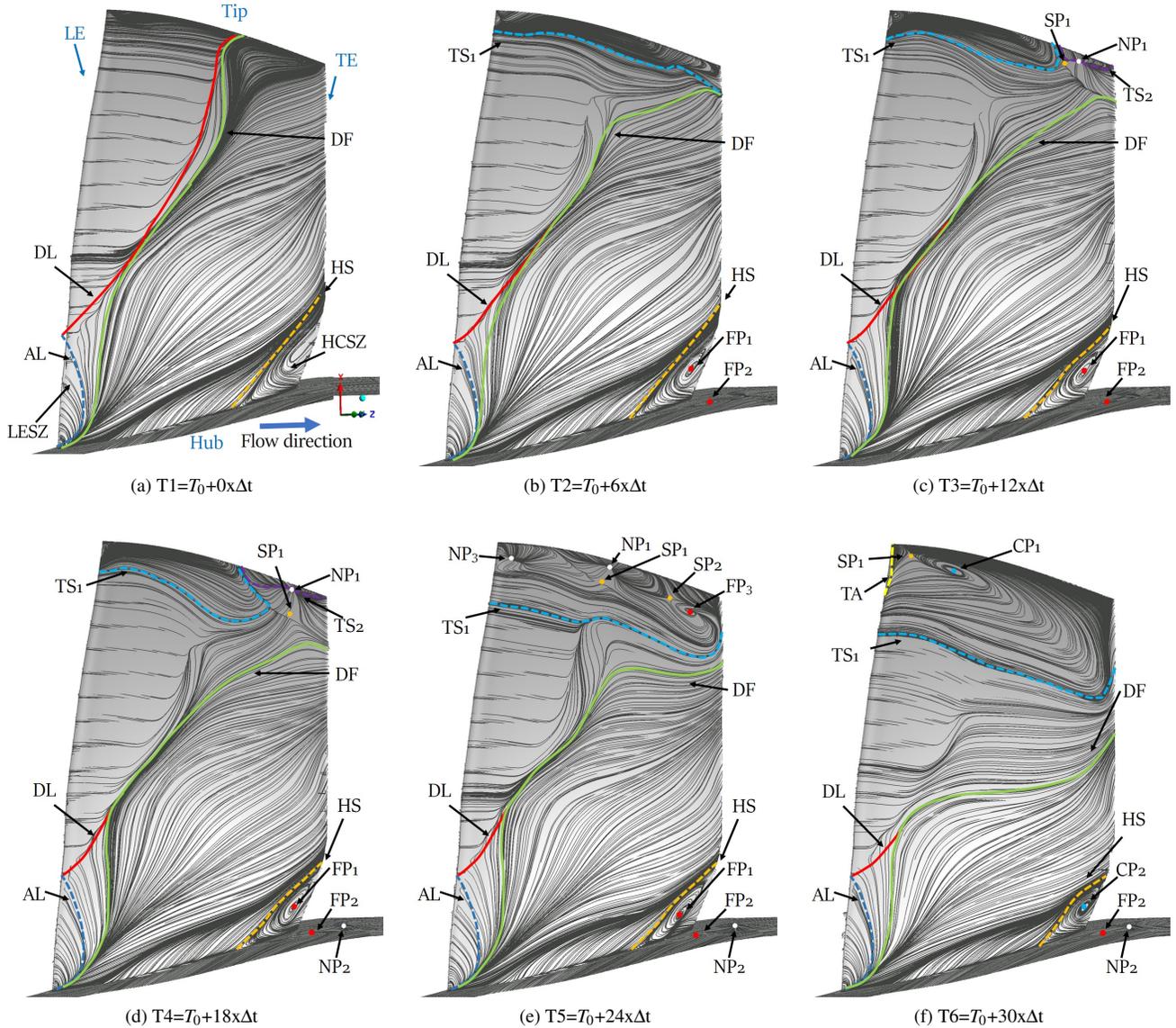

(a) T1=$T_0$+0x$\Delta$t

(b) T2=$T_0$+6x$\Delta$t

(c) T3=$T_0$+12x$\Delta$t

(d) T4=$T_0$+18x$\Delta$t

(e) T5=$T_0$+24x$\Delta$t

(f) T6=$T_0$+30x$\Delta$t

FIG. 12: Evolution of suction surface streamlines at various time instances from the near-stall point for the rough case.

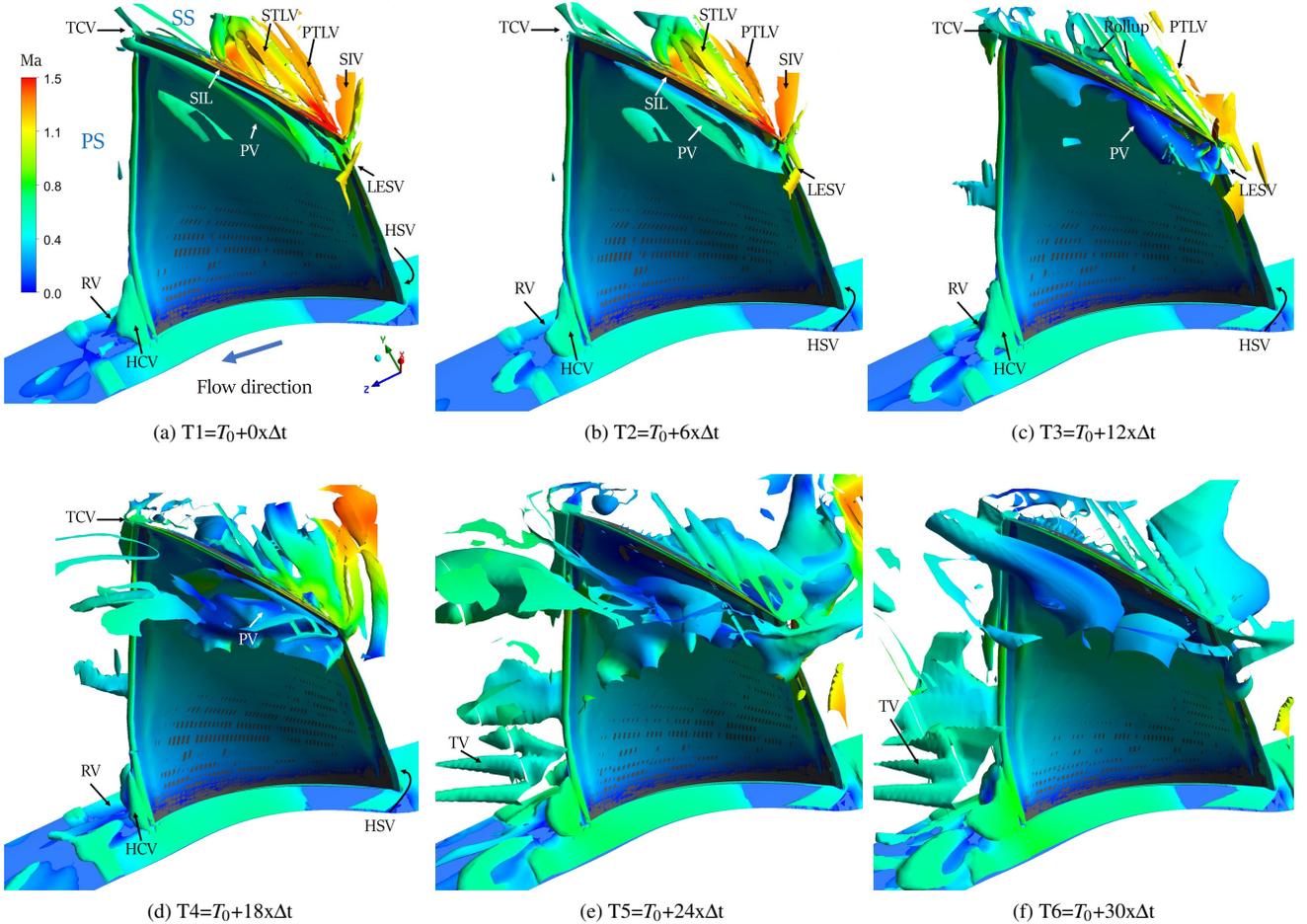

FIG. 13: Evolution of various secondary flow structures shown by the Q-criterion (Q = $1.35 \times 10^{-4}$) and superimposed by Mach number contours at different time instances for the smooth case.

## C. Visualisation of Secondary flow Structures using Q-criterion

Numerical visualization utilizing the Q-criterion is a vital tool for comprehensively understanding complex 3D flow fields. The Q-criterion provides a quantitative assessment of local fluid element rotation and deformation, offering invaluable insights into various intricate flow features such as vortices, shear layers, wakes, and mixing layers. It is particularly adept at characterizing these 3D complexities and is defined by the expression where $\Omega_{ij}$ and $S_{ij}$ represent the symmetric (rotation rate tensor) and asymmetric (strain rate tensor) components of the velocity gradient tensor, respectively.

$$Q = \|\Omega_{ij}\|^2 - \|S_{ij}\|^2,$$ (3)

$$\Omega_{ij} = \frac{1}{2}\left(\frac{\partial u_i}{\partial x_j} + \frac{\partial u_j}{\partial x_i}\right),$$ (4)

$$S_{ij} = \frac{1}{2}\left(\frac{\partial u_i}{\partial x_j} - \frac{\partial u_j}{\partial x_i}\right).$$ (5)

Figure 13 depicts instantaneous plots of various turbulent structures in the rotor passage using the Q-criterion (Q=$10^{-4}$) superimposed by Mach number contours viewed from the pressure side. At $T_1$, the main vortex structures near the hub consist of the horseshoe vortex (HSV), formed by the flow moving around the blade's LE (not visible); the ramp vortex (RV), induced by the shape of the hub surface; and the hub corner vortex (HCV), generated by flow separation at the suction side hub corner zone. There is a shockwave at the tip LE which interacts with the boundary layer and forms a shock-induced vortex (SIV). The LE shock impingement location (SIL) is shown. The tip region also has primary and secondary tip leakage vortices (PTLV and STLV) that are formed due to the pressure difference between the pressure side and the suction side of the rotor blades. The flow at the tip of the blade spills over from the pressure surface to the low-pressure suction surface, creating a swirling motion in the form of PTLV and STLVs. There is a tip corner vortex (TCV), which originates from the TE tip. A LE spillage vortex (LESV) is also present. There is vortex shedding in the blade passage between the suction and pressure sides. As time progresses these blade passage vortices grow in size due to the fluctuating pressure gradients. Up to $T_4$, the vortices in the hub region do not grow and maintain a similar size as $T_1$. At $T_2$ most of the vortex structures in the tip region maintain their size, shape, and strength. The LE shockwave starts to reduce in its strength and the portion of supersonic flow over the blade reduces due to the increasing streamwise APG. This causes the SIV to weaken and then disappear in $T_3$. At $T_3$, the increasing APG, causes the flow to separate from the LE, which makes the LE shock vanish. The PTLV go into a roll-up. The TCV starts to gain strength due to the increased flow separation from the tip region. At $T_4$, the PTLV begins to lose its distinctness due to the reducing pressure gradient across the pressure and suction sides of the blade. At $T_5$ and $T_6$, the flow becomes quite random and chaotic due to the extreme flow separation. There are large rotor-stator passage vortex sheets that form. The hub corner vortices give way to tornado vortices (TV) due to the increased pressure gradients in the hub region and due to increased instability from the formation of foci in the hub corner region (Fig. 11f). All these vortices result in massive flow blockage which results in rotor stall.

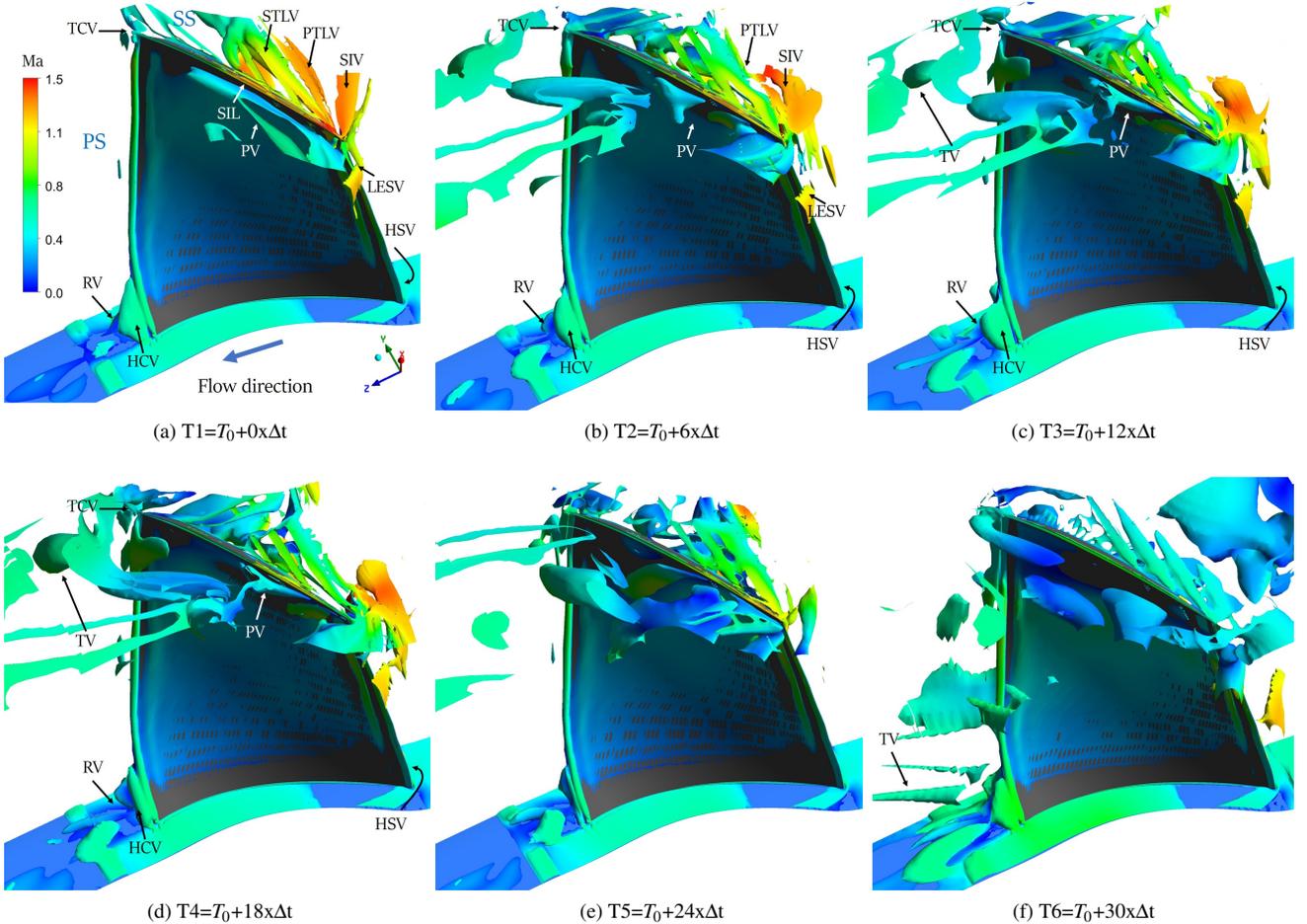

FIG. 14: Evolution of various secondary flow structures shown by the Q-criterion (Q = $1.35 \times 10^{-4}$) and superimposed by Mach number contours at different time instances for the rough case.

The Q-criterion (Q = $1.35 \times 10^{-4}$) superimposed with Mach number contours is also shown for the rough case in Figure 14. The vortex structures at $T_1$ are quite similar to that of the smooth case, except that there is a less-pronounced SIV due to a slightly lower Mach number at the LE. The TCV is larger due to the increased mixing of flow due to the rough boundary layer. At $T_2$, due to the onset of flow separation across the entire blade, the vortex structures begin to get more violent and chaotic. The SIV and STLV, LESV begin to vanish and the TCV becomes larger. A passage vortex sheet begins to form in the rotor-stator passage. The PTLV roll-up is disrupted in the rough case. At $T_3$, increased vortex shedding is observed in the rotor blade passage and there is almost no distinct PTLV. TVs begin to form downstream of the tip TE, which were not seen for the smooth case and their increased growth is visible in $T_4$. From $T_5$ to $T_6$ the various vortices break down further which causes an increase in the blockage, which is greater in the rough case, eventually leading to stall.

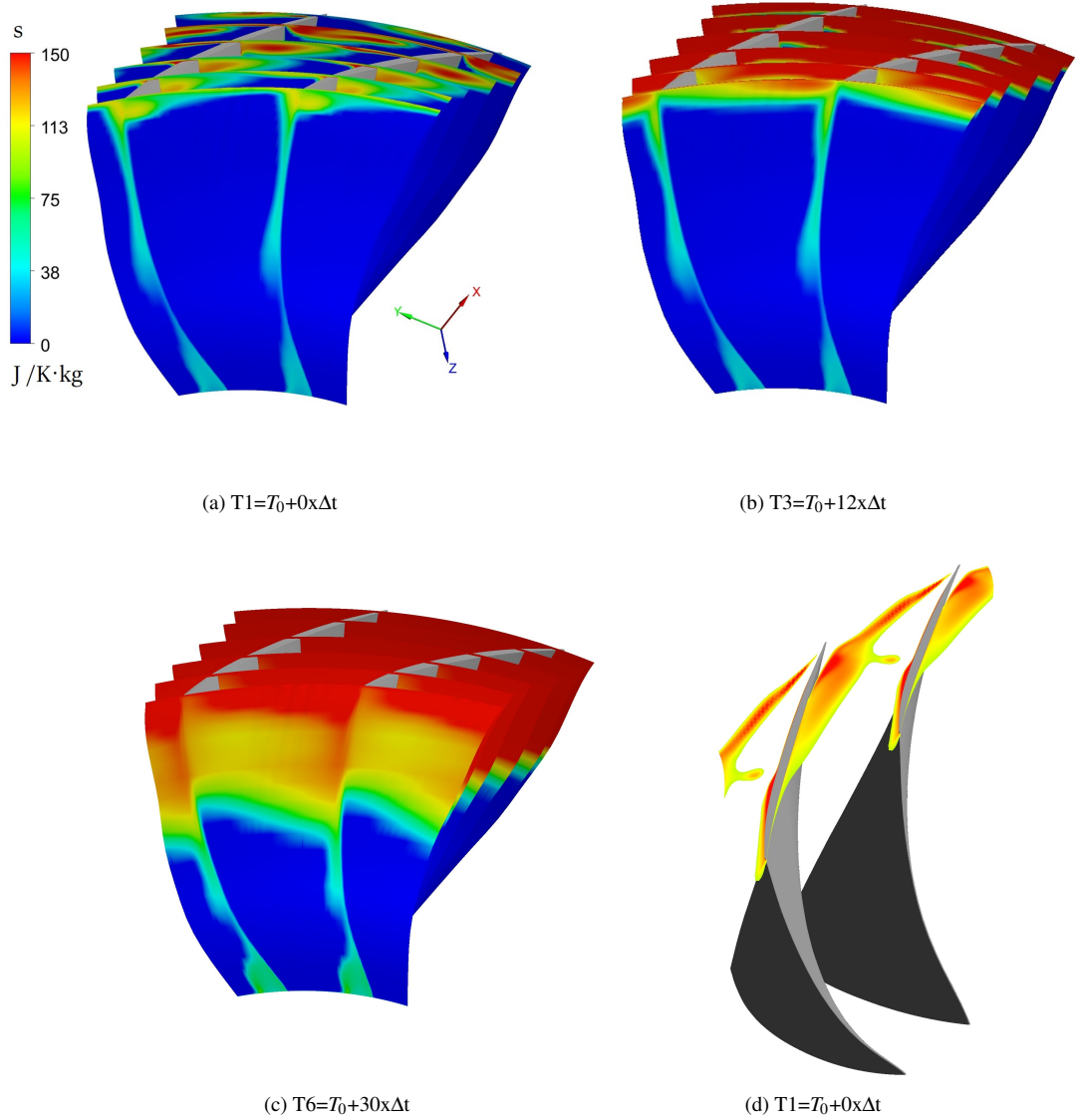

(a) T1=$T_0$+0x$\Delta$t

(b) T3=$T_0$+12x$\Delta$t

(c) T6=$T_0$+30x$\Delta$t

(d) T1=$T_0$+0x$\Delta$t

FIG. 15: (a-c) Spanwise entropy contours for various time instances and (d) iso-clips of entropy in the rotor passage (z = 98.75%) at $T_1$ for the smooth case.

### D. Entropy and Flow Blockage

Figures 15 and 16 show instantaneous spanwise entropy contours for 3-time instances (a-c) and iso-clips of entropy ($s \geq 100 \, J/K \cdot kg$) in the rotor passage (span (z) = 98.75%) at $T_1$ (d) for the smooth and rough cases, respectively. These sections are located at 0.05, 0.2, 0.4, 0.6, 0.8, and 1 fraction of the tip axial chord. At $T_1$ and at the blade TE, there is qualitative similarity between the smooth and rough cases at all spanwise locations, except the tip region. Quantitatively, there are significantly higher entropy levels in the rough case than in the smooth case, especially in the tip regions. At $T_1$, distinct TLVs are visible from the entropy plots (Figures 15 and 16 (d)). From Fig. 16 (d) it is evident that the TLV interacts with the thicker boundary layer and hence there is greater entropy generation. Due to interaction with the thicker rough boundary layer, there is increased flow mixing with the TLVs and thus they begin to vanish sooner with time than the smooth case. At $T_3$, the flow begins to mix out further leading to larger losses in both cases. For the rough case, there are high entropy levels from 85% span as compared to 93% span of the smooth case. At $T_6$, the flow in the entire tip region has become very chaotic and there is very high flow blockage due to the breakdown of vortices, which indicates imminent stall, especially in the rough case. Entropy and blockage in a compressor are related in that an increase in entropy typically signifies higher energy losses within the flow, which often correlates with increased blockage. As the blockage increases, the flow is constricted, leading to higher dissipation of energy and, consequently, a rise in entropy. This increase in entropy indicates degraded aerodynamic performance and efficiency of the compressor. The change in blockage affects the meridional velocity and hence the loading on the blade surface which makes it imperative to quantify the blockage. The blockage in a rotor passage is estimated by a blockage parameter $B$, proposed by Suder.[51] The blockage parameter is defined as the ratio of the area lost due to blockage to the available geometric area, as shown below.

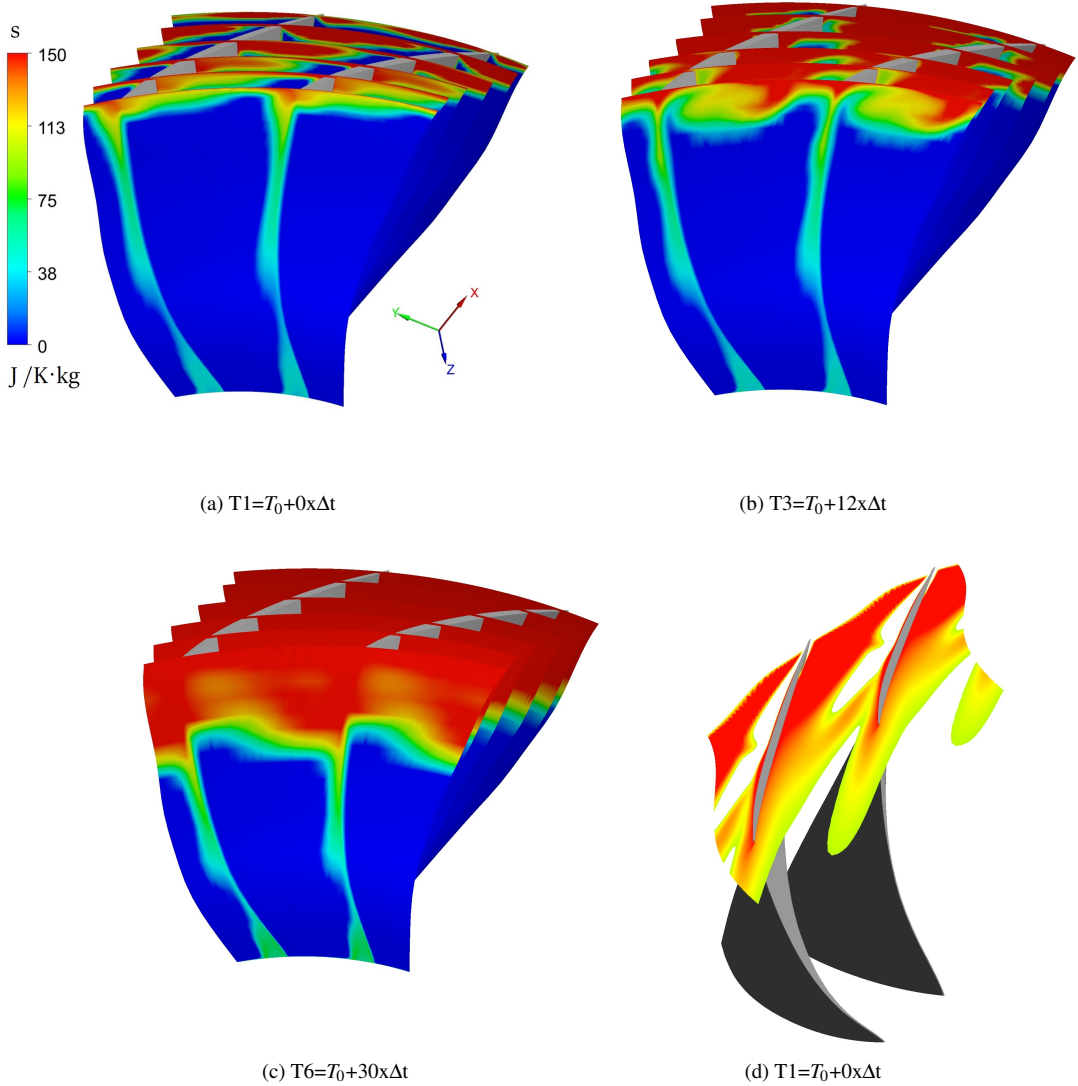

(a) T1=$T_0$+0x$\Delta$t

(b) T3=$T_0$+12x$\Delta$t

(c) T6=$T_0$+30x$\Delta$t

(d) T1=$T_0$+0x$\Delta$t

FIG. 16: (a-c) Spanwise entropy contours for various time instances and (d) iso-clips of entropy in the rotor passage (z = 98.75%) for the rough case at $T_1$.

$$B = 1 - \frac{A_{flow}}{A_{geom}}, \qquad (6)$$

where $A_{flow}$ is the effective flow area of the rotor having a geometric area $A_{geom}$. For an axial flow machine, $A_{flow}$ is calculated as the ratio of actual mass flow (product of density and axial velocity) to the mass flow if the flow was inviscid.[45] Along a constant radius line, the blockage parameter can be

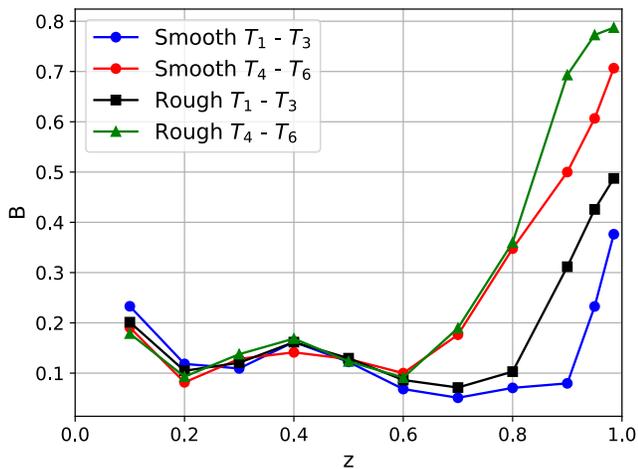

FIG. 17: Blockage parameter (B) as a function of the spanwise fraction (z) for smooth and rough cases.

expressed as

$$B = 1 - \frac{(\rho u_{axial})_{av}}{(\rho u_{axial})_{invisc}}, \qquad (7)$$

where $(\rho u_{axial})_{av}$ represents the average axial momentum along the constant radius line (span) and $(\rho u_{axial})_{invisc}$ is the average inviscid axial momentum along the constant radius line.

Figure 17 shows the variation in the blockage parameter from hub to shroud plotted against the fraction of span (z), at the TE surface. There are two lines each for the smooth and rough cases. One line for each case represents the average blockage at each spanwise location from $T_1$ to $T_3$ and the other line is the average blockage from $T_4$ to $T_6$. For all the cases, there is about 18-22% blockage at 10% span due to the hub corner separation. At 20% span, the effect of the hub corner separation in the blockage parameter reduces. Subsequently, there is again an increase in blockage due to the profile losses in the midspan section of the blade. The blockage then begins to drop till about 60-70% span. From 60% span pronounced differences are visible between the rough and smooth cases for both time scenarios. From $T_1$ to $T_3$, there is a significant rise in blockage after 80% span, whereas, for the smooth case, this rise is visible after 90% span. These differences in the blockage in the tip region are visible from the entropy plots in Figs. 15 and 16. The average blockage from 90% span to the tip from $T_1$ to $T_3$ for the smooth and rough cases are 21.7% and 40.5% respectively. For the time period between $T_4$ to $T_6$, there is significantly higher blockage for both cases, due to the chaotic flow field and higher losses. The flow blockage begins to increase from 60% span. Beyond 80% span, the rough case offers higher blockage than the smooth case due to increased flow mixing. The average blockage from 90% span to the tip from $T_4$ to $T_6$ for the smooth and rough cases are 59.6% and 75.2% respectively. Thus when the flow approaches a complete stall, there is immense blockage of flow in the tip region, all the more for the rough case.

## IV. CONCLUSIONS

This study investigates the flow physics at near-stall conditions using steady and unsteady RANS calculations for smooth and rough NASA rotor 67 blades. The RANS data is extensively validated against experiments and the numerical results show good agreement. The near-stall flow physics within the rotor-stator passages is thoroughly analyzed to comprehend the evolution of various secondary flow structures and unsteadiness for both smooth and 30 $\mu m$ rough cases, which is the average surface roughness developed on an operational blade after approximately 20,000 cycles. From RANS calculations, it is evident that roughness reduces the blade's performance. The 2 $\mu m$ rough case showcases flow characteristics which are quantitatively and all the more qualitatively similar to the smooth case, except close to the shroud. The 30 $\mu m$ rough case has much higher TKE and entropy in its boundary due to greater flow mixing due to the rough surface, resulting in greater losses. There is higher loss especially in the tip region due to the formation of tip leakage vortices.

Unsteady RANS calculations show that due to boundary layer roughness, the shock impingement location for the 30 $\mu m$ rough case is slightly upstream than the smooth case. This shock propagates radially downwards and creates flow deflection lines and a hub leading edge separation zone for both smooth and rough cases. There is also the formation of a hub corner separation bubble. The topology of the flow in the hub region for the smooth and rough cases is quite similar at near-stall. While approaching the complete stall, the rough case undergoes far greater instability than the smooth case. The flow separates across the tip region much earlier than the smooth case and also alternates frequently between one or two separation regions due to the rough boundary layer. The number of nodes, indicating greater flow instability, in the tip region is greater in the rough case than in the smooth case. At the fully stalled condition, there are large circulation zones in both the rough and smooth cases, with the center for the smooth case being much closer to the LE. The extent of the tip separation is up to 78% span, whereas it is up to 70% in the smooth case. The rough boundary layer reduces the extent of the separation zone from the tip.

The rough boundary layer triggers the early breakdown of the distinct vortex structures. The breakdown of various vortices, for both cases, leads to increased flow mixing and eventually massive blockage at the fully stalled condition. Entropy contours at the TE show, greater losses in the tip region for the rough case than the smooth case as the blockage from the rough case is significantly higher than the smooth case in the tip region. From the onset of stall to the fully stalled conditions, the blockage from 90% span to the tip varies from 21.7% to 59.6% and from 40.5% to 75.2% in the smooth and rough cases respectively. This massive blockage, due to the breakdown of vortices and the chaotic flow structures, leads to a complete stall. The understanding gained from the qualitative and quantitative analysis of smooth and rough NASA rotor 67 blades near-stall would be helpful in the compressor and fan blade design exercise.

## ACKNOWLEDGEMENTS

Harshal Akolekar acknowledges the seed grant support from IIT Jodhpur (I/SEED/HDA/20230206).

## AUTHOR DECLARATIONS

**Conflict of Interest**
The authors have no conflicts to disclose.

**Author Contributions Prashant Godse:** Conceptualization (equal); Data curation (lead); Formal analysis (lead); Visualization (lead); Writing - original draft (supporting). **Harshal Akolekar:** Conceptualization (equal); Data curation (supporting); Investigation (supporting); Supervision (equal); Formal analysis (supporting); Visualization (supporting); Writing original draft (lead); Writing review and editing (lead). **AM Pradeep:** Conceptualization (equal), Supervision (equal); Formal analysis (supporting); Writing original draft (supporting); Writing review and editing (supporting).

## NOMENCLATURE

| | |
|---|---|
| AL | Attachment line |
| APG | Adverse pressure gradient |
| B | Blockage parameter |
| CFD | Computational Fluid Dynamics |
| CP | Circulation point |
| DF | Hub drifted flow line |
| DL | Flow deflection line |
| FP | Focal point |
| HCSZ | Hub corner separation zone |
| HCV | Hub corner vortex |
| HS | Hub separation line |
| HSV | Horseshoe vortex |
| LE | Leading edge |
| LESV | Leading edge spillage vortex |
| LESZ | Leading edge separation zone |
| NASA | National Aeronautics and Space Administration |
| NP | Nodal point |
| PS | Pressure side |
| PTLV | Primary tip leakage vortex |
| PV | Passage vortex |
| Q | Magnitude of Q-criterion |
| RANS | Reynolds Averaged Navier Stokes |
| RMS | Root mean squared |
| RV | Ramp vortex |
| SIL | Shock impingement location |
| SIV | Shock induced vortex |
| SP | Saddle point |
| SS | Suction side |
| STLV | Secondary tip leakage vortex |
| TA | Tip attachment line |
| TCV | Tip corner vortex |
| TE | Trailing edge |
| TKE | Turbulent kinetic energy |
| TLF | Tip leakage flow |
| TLV | Tip leakage vortex |
| TV | Tornado vortex |
| URANS | Unsteady RANS |
| $A_{flow}$ | Effective flow area |
| $A_{geom}$ | Geometric area |
| k | Turbulent kinetic energy |
| Ma | Mach number |
| Pa | Pascal |
| $s$ | Entropy |
| $S_{ij}$ | Strain rate tensor |
| $\Delta t$ | Time interval |
| T | Time |
| $u_{axial}$ | Axial velocity |
| z | Span fraction |
| $\eta$ | Isentropic efficiency |
| $\tau$ | Wall shear stress magnitude |
| $\mu$ | Dynamic viscosity |
| $\mu_t$ | Eddy viscosity |
| $\Pi$ | Pressure ratio |
| $\phi$ | Normalized mass flow rate |
| $\rho$ | Density |
| $\omega$ | Specific dissipation rate |
| $\Omega_{ij}$ | Rotation rate tensor |

## DATA AVAILABILITY

The data that support the findings of this study are available from the corresponding author upon reasonable request.

Western Reserve University, 1996).